\documentclass[aps,prx,reprint,superscriptaddress]{revtex4-2}

\usepackage[T1]{fontenc}
\usepackage[utf8]{inputenc}
\usepackage{amsmath,amssymb}
\usepackage{graphicx}
\usepackage{booktabs}
\usepackage{hyperref}
\usepackage{url}
\hypersetup{colorlinks=true,linkcolor=blue,citecolor=blue,urlcolor=blue}

\begin{document}

\title{Earthquake Dynamics as a Benchmark for Scale-Invariant Behavior}

\author{Kilian Duplat}
\affiliation{Universit\'e Lyon 1, CNRS, Institut Lumi\`ere Mati\`ere, UMR5306, F-69622 Villeurbanne, France.}
\author{German Varas}
\affiliation{Instituto de F\'isica, Pontificia Universidad Cat\'olica de Valparaiso (PUCV), Avenida Universidad 330, Valparaiso, Chile.}
\author{Osvanny Ramos}
\email{osvanny.ramos@univ-lyon1.fr}
\affiliation{Universit\'e Lyon 1, CNRS, Institut Lumi\`ere Mati\`ere, UMR5306, F-69622 Villeurbanne, France.}

\begin{abstract}
The fractal dimension of a structure can be directly compared with Euclidean spatial dimensions. Its counterpart in scale-invariant dynamics is the exponent $\tau$ in the power-law distribution of event sizes, \(P(s)\sim {s}^{-\tau{}}\). Yet no reference value exists for comparing $\tau$ across systems, and consequently no universal baseline for characterizing the associated dynamics. Here, we propose earthquake dynamics, characterized by $\tau$ $\simeq$ 1.67, as a benchmark for scale-invariant behavior. Analyses of Southern California and Japan catalogs yield energy distributions with $\tau$ $\simeq$ 1.67. Using synthetic distributions generated over a wide range of $\tau$, we show that changing $\tau$ leads to qualitatively different dynamics, not just quantitative variations. Earthquake-like behavior is limited to 1.5 $\leq$ $\tau$ < 2.0, and narrows to 1.58 $\leq$ $\tau$ $\leq$ 1.76 when energy variations are constrained to within a factor of ten of real earthquakes. For $\tau$ < 1.5, externally supplied energy is required to sustain a slowly driven fault, while small events dominate the energy budget for $\tau$ > 2.
\end{abstract}

\maketitle

\section{Introduction}
In many dissipative phenomena, including earthquakes \cite{ref1}, granular faults \cite{ref2,ref3,ref4,ref5,ref6,ref7}, sandpiles \cite{ref8,ref9,ref10} and subcritical rupture \cite{ref11,ref12,ref13,ref14,ref15,ref16,ref17}, energy is slowly accumulated and then released through sudden events of all sizes, typically following power-law distributions. The scale-invariant nature of these events motivated theoreticians to draw on the formalism of phase transitions \cite{ref18,ref19}. Yet this raised a fundamental question: in natural systems, how is the fine-tuning of the order parameter --required to reach criticality-- achieved? \cite{ref18} The idea of a critical point acting as an attractor of the dynamics, introduced by the Self-Organized Criticality (SOC) in 1987 \cite{ref19}, offered a compelling and elegant answer.

Although SOC's ambitious claims \cite{ref20} and the absence of a unified theoretical framework attracted criticism \cite{ref21,ref22}, the conceptual power of the idea galvanized leading figures in statistical physics \cite{ref23,ref24} and propelled its application to fields as diverse as seismology \cite{ref25}, neuroscience \cite{ref26}, and even financial markets \cite{ref27}.

Earthquakes were the most familiar, well-studied, and arguably the most relevant of these phenomena, and thus quickly became the reference point for interpreting scale-invariant behavior. Regardless of the value of the power-law exponent $\tau$ in the event-size distribution \(P(s)\sim {s}^{-\tau{}}\), the underlying interpretation remained essentially the same: events occur across all scales, with numerous small ones and rare, catastrophic ones that dominate the total energy release.

The central message of this article is that this general interpretation can be misleading. Varying the exponent $\tau$ may lead to qualitatively different scenarios, rather than merely quantitative variations. In particular, systems characterized by different $\tau$ values may exhibit dynamics that differ fundamentally from those observed in real seismicity.

How did this general interpretation become established? In the 1990s, most laboratory experiments and earthquake catalogs lacked the precision required to critically test or guide theoretical developments. Reported b-values in the Gutenberg--Richter (GR) law spanned a wide range \cite{ref28}, and no clear consensus existed on how to define an avalanche in a way that enabled meaningful comparison between theoretical or simulated $\tau$ exponents and those inferred from real data. From a theoretical perspective, much of the effort focused on determining the value of $\tau$ and classifying avalanches into universality classes \cite{ref29,ref30,ref31,ref32,ref33}. However, the extent to which the underlying dynamics differ across these classes was rarely examined, and they were often implicitly assumed to exhibit the same earthquake-like behavior described above.

If we draw an analogy between scale-invariant (fractal) structures and scale-invariant dynamics, the exponent $\tau$ plays a role analogous to that of the fractal dimension. This concept, introduced by Mandelbrot in the 1970s \cite{ref34}, is rather intuitive. By comparison with Euclidean space dimensions, one can assess how closely a fractal structure resembles isolated points (\textit{d = 0}), a line (\textit{d = 1}), a plane (\textit{d = 2}), or a fully three-dimensional object (\textit{d = 3}). However, there is currently no standard reference value against which $\tau$ can be compared, and consequently no universal baseline for characterizing the associated dynamics.

Here, we propose earthquake dynamics, characterized by $\tau$ $\simeq$ 1.67 \cite{ref35}, as a natural reference point for assessing scale-invariant behavior. By doing so, we challenge the common interpretation that directly associates scale-invariant dynamics with earthquake-like behavior.

\section{Avalanche definition}
A central goal of this article is to clarify the physical scenarios associated with particular exponent values. As expected, however, different definitions of avalanche size lead to distinct event distributions. Consider a power law of the form \(P(s)=\frac{1}{N}{s}^{-{\tau{}}_{1}}\), where \textit{N} is a normalization constant. The variable \textit{s} can be expressed as \(s={{s}_{l}}^{{D}_{A}}\), where \({s}_{l}\) is the linear extent of the avalanche and \({D}_{A}\) its fractal dimension. Using this relation we obtain:

\begin{equation}
P(s)\,ds=P(s_l)\,ds_l
\end{equation}
\begin{equation}
\frac{1}{N}s_l^{-\tau_1 D_A}D_A s_l^{D_A-1}\,ds_l=P(s_l)\,ds_l
\end{equation}
where
\begin{equation}
\tau_2=(\tau_1-1)D_A+1
\end{equation}
Thus, the same underlying process can be described by two power laws, \(P(s)\) and \(P({s}_{l})\), characterized by different exponents, $\tau$\textsubscript{1} and $\tau$\textsubscript{2}. Which definition should be considered the "correct" one?

For elastic interfaces, where exponent values have been analytically calculated for different universality classes \cite{ref33}, \textit{avalanche}\textit{ size is conventionally defined as the} \textit{displacement integrated over the interface}, \(S=\int {d}^{n}x\,\Delta u(x)\), where \textit{n} is the internal dimension of the interface and \(\Delta u(x)\) is its local displacement during the event. This definition incorporates both the spatial extent of the avalanche and the displacement within it, providing a reference for comparing avalanche statistics across systems.

For earthquakes, the corresponding observable is the scalar seismic moment, \({M}_{0}=\mu{}\int dA\,D(x)=\,\mu{}A\overline{D}\), for uniform shear modulus \(\mu{}\) and a common slip direction \cite{ref36}. Here, \textit{A} is the rupture area and \(\overline{D}\) is the average relative displacement across the fault. Although slip is several orders of magnitude smaller than the rupture’s linear dimensions, both rupture area and average slip vary over broad ranges and jointly contribute to seismic-moment scaling \cite{ref37,ref38}. Seismic moment is therefore proportional to the integrated slip over a two-dimensional fault surface, providing the earthquake counterpart of the avalanche-size definition used for elastic interfaces \cite{ref33}.

\section{Earthquakes Statistics}
Catalogs report earthquake magnitudes M, typically derived directly from seismic moments, as a measure of event strength. From these magnitudes, the released energy E can be estimated using \cite{ref39}:

\begin{equation}
\mathrm{Log}(E)=\frac{3}{2}(M+6.07).
\end{equation}
We analyze two local catalogs from Japan and California, which are among the most extensively studied seismic regions (Fig. 1):

\begin{itemize}
  \item JMA catalog from the Japan Meteorological Agency \cite{ref40}; Time interval: [1983-2023); Region: Longitude 123\textsuperscript{0}E -- 163\textsuperscript{0}E, Latitude 26\textsuperscript{0}N -- 50\textsuperscript{0}N.
  \item SCEDC catalog from the Southern California Earthquake Data Center \cite{ref41}; Time interval: [1980-2025); Region: Longitude 113.1\textsuperscript{0}W -- 121.7\textsuperscript{0}W, Latitude 30.1\textsuperscript{0}N -- 39.3\textsuperscript{0}N.
\end{itemize}
Most recent geophysical studies analyze the GR law locally, in both space and time (see an exhaustive analysis here \cite{ref42}). Local variations may arise from statistical artifacts or from genuine changes in the physical conditions of the fault. Discrepancies among different earthquake catalogs are also a recognized issue, reflecting the difficulties in accurately measuring earthquake magnitudes \cite{ref42}.

To minimize these issues, we focus on the \textit{primary archives} for the two selected regions and perform a global analysis. First, we analyze the annual number of earthquakes for all different magnitude values. At smaller magnitudes, one observes clear global variations, which are attributed to changes in detection capabilities (e.g., new sensors or improved procedures). The data analysis begins with a completeness test, in which the minimum magnitude considered reliable for fitting is defined as the lowest value showing a stable detection rate over the selected time interval, corresponding to Magnitude M2$\equiv$[M2.0, M2.9] in California and M5$\equiv$[M5.0, M5.9] in Japan. Using the same criteria, we have reduced the temporal window of analysis in Japan to the last 23 years.

\begin{figure*}[t]
  \centering
  \includegraphics[width=0.6\textwidth]{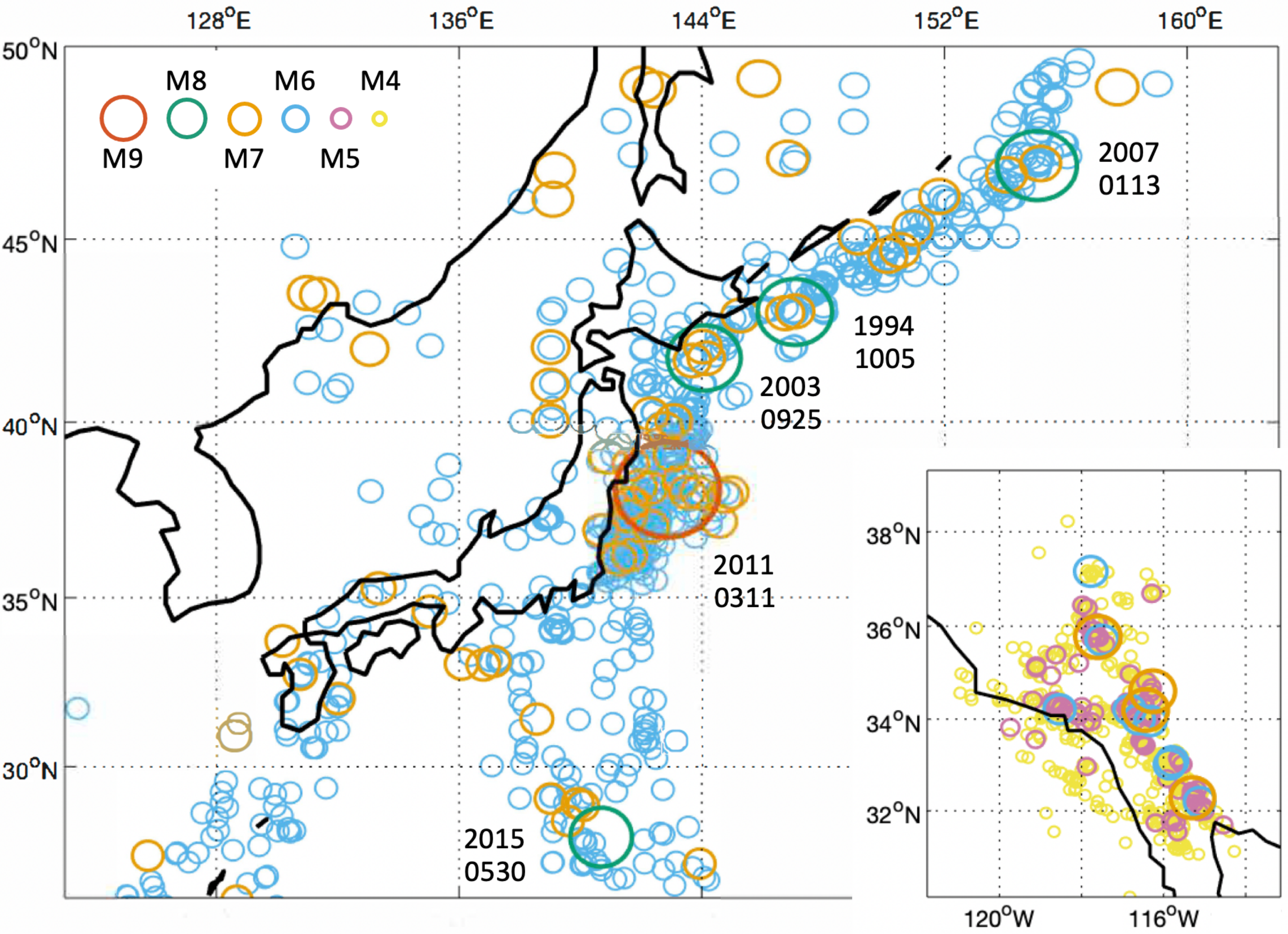}
  \caption{\textbf{Catalogs}. Regions analyzed in the JMA (Japan) and SCEDC (Southern California) catalogs in the time-intervals reported in the text. All M6-M9 earthquakes in Japan and all M4-M7 earthquakes in Southern California are shown, with colors and symbol sizes indicating magnitude (see legend). The dates of the Tōhoku earthquake (M9.1) and the four M8 events in Japan are also indicated. The M8 earthquakes in Japan are used as reference events throughout the article.}
  \label{fig:1}
\end{figure*}
The GR law is defined as:

\begin{equation}
\mathrm{Log}(N_c(M))=a-bM,
\end{equation}
where \textit{a} and \textit{b} are constants and N\textsubscript{c}(M) is the (cumulative) number of earthquakes of magnitude M or greater. The GR relations for both regions, with N\textsubscript{c}(M) expressed on a yearly basis, yield \textit{b} values very close to 1.0 (Fig. 2a).

\begin{figure*}[t]
  \centering
  \includegraphics[width=0.98\textwidth]{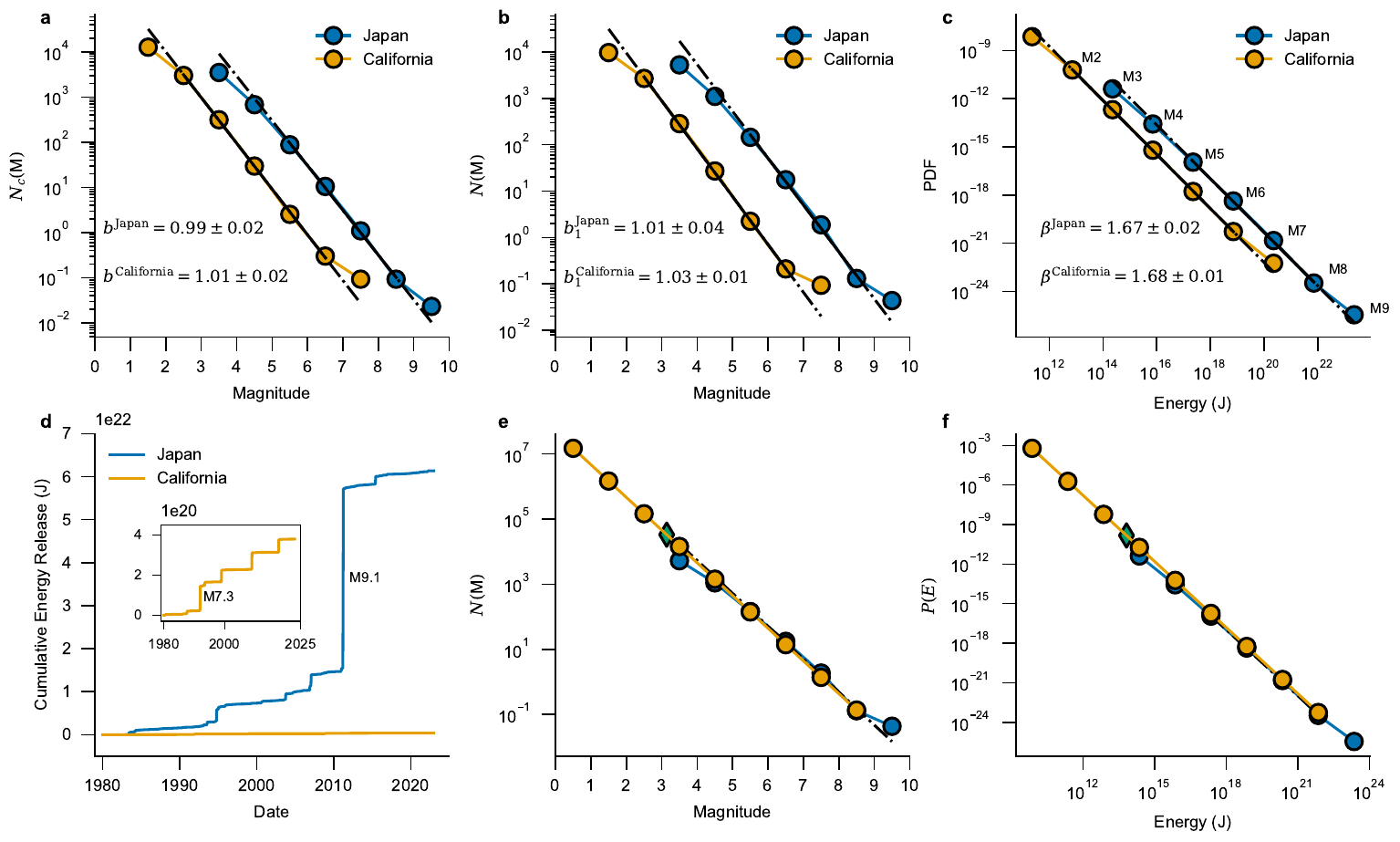}
  \caption{\textbf{Earthquakes statistics.} (a) Gutenberg-Richter relations (Eq. 5) for Japan (JMA catalog) and Southern California (SCEDC catalog). The fits correspond to the solid lines. Magnitudes smaller than M2$\equiv$[M2.0, M2.9] in the SCEDC, and than M5 in JMA, are affected by catalog incompleteness. (b) Non-cumulative number of earthquakes per year as a function of the magnitude (Eq. 6). (c) Non-cumulative energy distribution. They follow power law distributions with $\beta$ values around 1.67. (d) Cumulative energy release, using the whole time interval of the catalogs. Notice that the Tōhoku earthquake (2011, M9.1) accounts for more than 63 \% of the total energy release in the period. Californian earthquakes release two orders of magnitude less energy. Four M7+ quakes dominate the total energy release: Landers (1992, M7.3), Hector Mine (1999, M7.1), Baja California (2010, M7.2), and Ridgecrest (2019, M7.1). (e,f) Extrapolation of the relations obtained in panels (b) and (c) for Japan, assuming \textit{b}\textit{\textsubscript{1}}=1 and $\beta$=1.67 respectively. Green diamonds correspond to the mean value of the released energy $\langle$E$\rangle$.}
  \label{fig:2}
\end{figure*}
In most areas of physics, cumulative distributions are not the primary object of analysis, since they tend to obscure the physical mechanisms associated with events of a given size. Instead, differential (or noncumulative) distributions are usually preferred because they provide direct access to the statistics of events within a specific size range, allowing one to relate scaling exponents and distribution shapes more transparently to the underlying physical processes. In (Fig. 2b) we define N(M) as the number of earthquakes of magnitude M per year, using integer intervals of magnitude values. The relation is similar to the GR relation,

\begin{equation}
\mathrm{Log}(N(M))=a_1-b_1M,
\end{equation}
with, as expected, the \textit{b}\textit{\textsubscript{1}} values being larger than their corresponding \textit{b} values. The differences are negligible in this case; however, we will show in the next section that for smaller \textit{b} values these differences become significant.

The scale-invariance of earthquakes is reflected in their energy distributions. Earthquake energy release is calculated from magnitude values using Eq. 4. Energy distributions follow power laws in the form $P(E)\sim E^{-\beta}$, with exponent values around $\beta$=1.67 \cite{ref35}, specifically $\beta$=1.67 $\pm$ 0.02 for Japan and $\beta$=1.68 $\pm$ 0.01 for Californian earthquakes (Fig. 2c). The time series of the cumulative energy release (Fig. 2d) further demonstrates that the largest earthquakes strongly dominate the total energy budget: the Tōhoku earthquake (M9.1) in Japan and four M7+ in California.

Japan, in particular, is significantly more active, with on average one catastrophic M8 earthquake approximately every decade (about every 6.6 years in our analysis). This is a well-established reference point --even part of the broader scientific culture-- which is why we will use it as a benchmark to evaluate how far a given dynamics deviates from real earthquake statistics in Japan.

Figures 2e and f show extrapolations of the distributions of magnitudes and energies for Japan earthquakes. We have chosen for the analysis micro-earthquakes of M0.0 as a baseline (choosing a smaller minimum delivers similar results) and 8.9 as the maximum one.

We exclude the M9.1 Tōhoku earthquake (2011), along with its associated energy release, from the extrapolation. This choice is motivated by several considerations. First, M9 events have recurrence intervals on the order of a century, which is much longer than the time span covered by our analysis. Second, including M9 events would extend the magnitude range up to M9.9, exceeding that of the largest recorded earthquake (M9.5, Valdivia, Chile, 1960). Finally, our goal is to construct a reference distribution with a maximum magnitude of M8 events, in order to compare it directly with observed M8 earthquakes in Japan. Excluding the Tōhoku event does not affect the distribution of M8 events.

\begin{figure*}[t]
  \centering
  \includegraphics[width=0.98\textwidth]{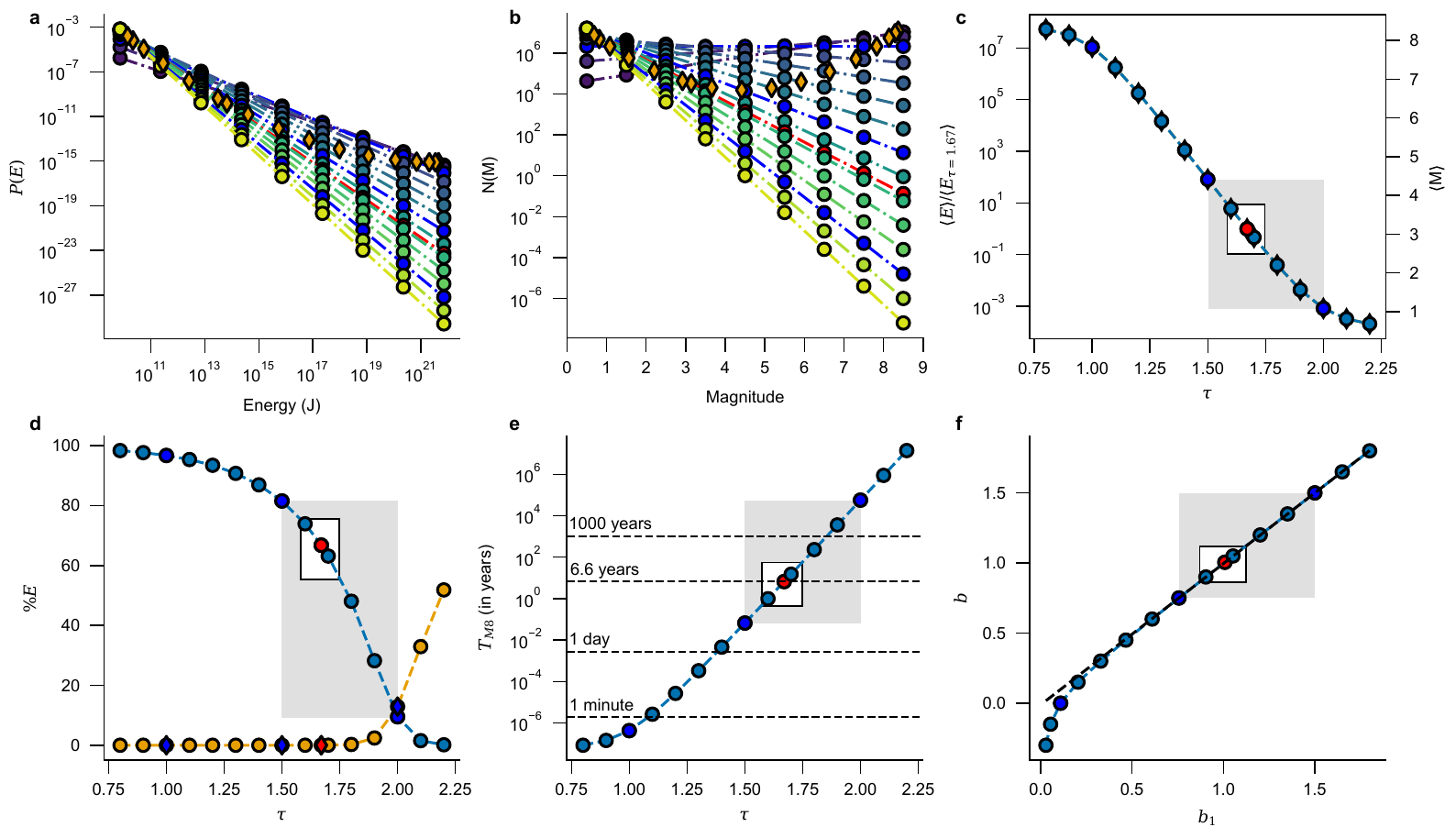}
  \caption{\textbf{Earthquake-like dynamics.} (a) Energy distributions for different exponent $\tau$ values ranging from 0.8 to 2.2 (dark to bright colors) considering a constant activity and magnitudes M$\in$[M0.0-M8.9]. In the case of $\tau$ = 1.67 (in red) it describes the energy distribution of Japan earthquakes. Curves with exponents $\tau$ = 1.0, 1.5 and 2.0 have been highlighted in blue. Orange diamonds correspond to the mean value of the released energy $\langle$ E$\rangle$. (b) The same data in (a) expressed as number of earthquakes / year as a function of the magnitude. (c) Average energy release $\langle$ E$\rangle$ normalized by the average energy release for $\tau$ = 1.67, and corresponding magnitude $\langle$ M $\rangle$ values as a function of the exponent. A gray rectangle indicates the earthquake-like regime, while the white rectangle indicates the range where the differences in $\langle$ E$\rangle$ are less than ten-times the one of real earthquakes. (d) Energy contributions (in \%) of extreme magnitude events M8 (in blue) and M0 (in orange) as a function of the exponent value. At $\tau$ =1.67, 67\% of the energy is released by the M8 events. However, for $\tau$ >2 the energy budget is controlled by the smallest M0 events. (e) Average period of M8 earthquakes as function of $\tau$. (f) b value as function of \textit{b}\textit{\textsubscript{1}}. The black dashed line has a slope equal 1. \textit{b} starts to deviate from the identity at \textit{b}\textit{\textsubscript{1}}<0.5, corresponding to $\tau$<1.33}
  \label{fig:3}
\end{figure*}
We have used \textit{b}\textit{\textsubscript{1}}=1 and $\beta$ = 1.67 for the extrapolation. Notice that from Eqs. 4 and 6, we obtain $\beta$=2\textit{b}\textit{\textsubscript{1}}/3+1. In the context of scale-invariant avalanches the size exponent is denoted $\tau$. Thus $\beta$=$\tau$=1.67 is the size exponent for earthquake dynamics.

The extrapolation provides the K value of the distribution $P(E)=K E^{-\tau}$. As $P(E)=N(E)\Delta E$, where N(E) is the annual number of earthquakes with a given energy, and $\Delta E$ the width of the logarithm binning, we can use

\begin{equation}
N_{\mathrm{tot}}=\sum P(E)\Delta E=\int P(E)\,dE
\end{equation}
\begin{equation}
E_{\mathrm{tot}}=\sum N(E)E=\int P(E)E\,dE
\end{equation}
to calculate the total annual energy release of E\textsubscript{tot}=1.621 $\times$ 10\textsuperscript{21} J and the total number of events of N\textsubscript{tot}=24 819 295 events/year, corresponding to an average activity rate of N\textsubscript{tot}\textsuperscript{-1}=1.27 s between events.

Considering a steady state, the input energy E\textsubscript{in}=$\langle$E$\rangle$ = E\textsubscript{tot}/N\textsubscript{tot}, which accounts for the energy stored in the fault in a time interval of 1.27 s. This corresponds to a mean magnitude value $\langle$M$\rangle$=M3.2, which is relatively small. The period of M8$\equiv$[M8.0, M8.9] events is T\textsubscript{M8}=6.6 years and they account for 67\% of the total energy release E\textsubscript{tot}.

Based on this information on real earthquakes in Japan, we can broadly define earthquake-like behavior as \textit{a regime characterized by a relatively small input energy rate E\_in leading to scale-invariant energy-release events, in which rare, extremely large events dominate the total energy budget} E\textsubscript{tot}.

\section{Changing exponent values}
By keeping the limits M0.0 and M8.9 and considering a constant activity (N\textsubscript{tot}=const.), we have calculated the energy distributions for different exponent values (Fig. 3a). At first glance we can think that all of them represent earthquake-like dynamics. However, if we use Eq. 4 and translate the energy values into magnitudes (Fig. 3b) it is much easier to interpret the data.

The range from 0 to 8.9 in magnitude corresponds to 13.5 decades in energy, which is enormous compared with the range reached by today's best experimental and numerical results. However, this "amplification" procedure of taking the exponent measured for a given dynamics and extrapolating it over the earthquake energy range, is precisely what enables a fair comparison with earthquakes, and is therefore essential for interpreting the dynamics in the light of earthquake statistics. Notice also that applying standard finite-size scaling to a given dynamics implicitly assumes that the scaling behavior remains valid in the limit N $\to$ $\infty$.

To assess the validity of our definition and its degree of similarity to real earthquakes across different exponent values, we focus on three criteria: (i) how small the input energy rate E\textsubscript{in} is (Fig. 3c); (ii) which events, and to what extent, control the total energy budget E\textsubscript{tot} (Fig. 3d); and (iii) how rare catastrophic M8 events are (Fig. 3e).

The dynamics for different values of $\tau$ span regimes ranging from extremely high mean energies $\langle$E$\rangle$, of the order of the energy of the largest events (below $\tau$=1.0), to extremely low ones, above $\tau$=2.0 (Figs. 3a,b,c). At $\tau$ = 1.5, where the $\langle$M$\rangle$=(M8.9-M0.0)/2 $\simeq$M4.5, the values of $\langle$E$\rangle$ split into two groups, corresponding to large and small energies (Fig. 3c). Consequently, we find \textbf{small} injected energies E\textsubscript{in} for $\tau$ $\geq$ 1.5.

However, it is important to note that the energy differences increase rapidly as $\tau$ moves away from $\tau$ = 1.67. In order to retain a range of values corresponding to scenarios close to those of real earthquakes, we arbitrarily restrict these differences to lie within one order of magnitude (i.e., ten-times) of the energy released by real seismicity. Under this criterion, the exponent values associated with dynamics comparable to those of real earthquakes are confined to the range 1.58 $\leq$ $\tau$ $\leq$ 1.76 (white rectangles in Fig. 3).

\begin{figure*}[t]
  \centering
  \includegraphics[width=0.98\textwidth]{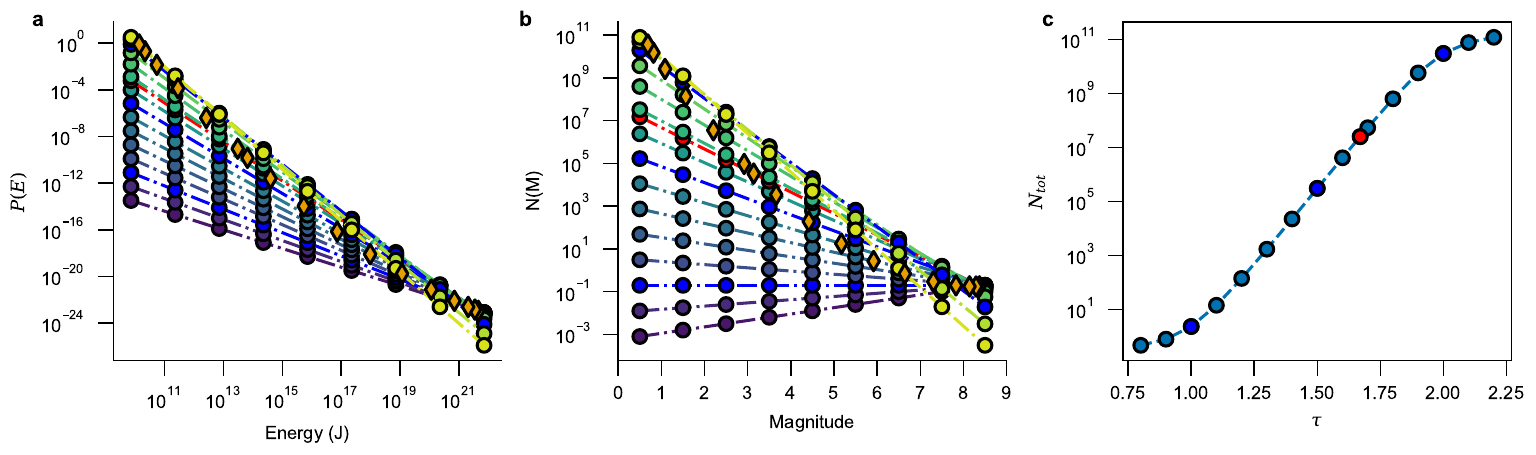}
  \caption{\textbf{Dynamics under constant energy}. (a) Energy distributions for different exponent $\tau$ values ranging from 0.8 to 2.2 (dark to bright colors) considering a constant energy release and magnitudes M$\in$[M0.0-M8.9]. In the case of $\tau$ = 1.67 (in red) it describes the energy distribution of Japan earthquakes. Curves with exponents $\tau$ = 1.0, 1.5 and 2.0 have been highlighted in blue. Orange diamonds correspond to the mean value of the released energy $\langle$E$\rangle$. (b) The same data in (a) expressed as number of earthquakes / year as a function of the magnitude. (c) Total number of events / year as a function of $\tau$.}
  \label{fig:4}
\end{figure*}
Figure 3d shows the percentage of energy released by the two extreme event classes, namely high-magnitude M8 and low-magnitude M0 events. The contribution of M8 events decreases rapidly for $\tau$ > 1.8. At $\tau$ = 2.0, all magnitudes contribute equally to the total released energy E\textsubscript{tot}, while for $\tau$ > 2.0 the smallest events dominate the energy budget. Using the ten-times limits defined in the previous analysis, the percentages of energy released by M8 earthquakes are found to be 75\% and 54\% (Fig. 3d), which are 12\% larger and 13\% smaller, respectively, than the energy release of real M8 earthquakes (67\%).

The combined analysis of Figs. 3c and 3d indicates that our broad definition of earthquake-like behavior is valid only in the regime 1.5 $\leq$ $\tau$ < 2.0, indicated by gray rectangles in Fig. 3.

The average period of M8 earthquakes under different exponent values (Fig. 3e) gives a clear indication of how qualitatively different the scenarios become when we move away from $\tau$=1.67 where T\textsubscript{M8}=6.6 years. Let us analyze the following scenarios:

$\tau$ =1.0: In this case all magnitudes are equiprobable with events of each of them happening in average every 13 seconds. In order to supply the energy for generating M8 earthquakes every 13 s in average in Japan, $\langle$E$\rangle$ is 1.1 $\times$ 10\textsuperscript{6} times larger than in the real earthquakes case, and corresponds to a M7.8 event. The scenario is quite far from the real earthquakes one. The energy cannot be slowly stored in the fault and predominantly liberated by rare extreme events. In this case the energy has to be always available to deliver an imminent extreme event, a behavior that can be found in some fracture experiments\cite{ref16}.

$\tau$ =1.5: With this exponent value, M8 earthquakes will take place every 24 days in average in Japan. $\langle$E$\rangle$ is 82 times larger than in the real earthquakes case, and corresponds to a M4.5.

$\tau$ =1.9: In this case, M8 earthquakes will take place every 3611 years in average in Japan. $\langle$E$\rangle$ is 232 times smaller than in the real earthquakes case, and corresponds to a M1.6 event. M8 events account for 28\% of the energy release in the fault.

$\tau$ =2.0: In this case, M8 earthquakes will take place every 56 490 years in average in Japan. $\langle$E$\rangle$ is 1204 times smaller than in the real earthquakes case, and corresponds to a M1.1 event. All magnitudes contribute equally to the energy budget. For larger exponent values M0 magnitude events control the energy budget. This is an issue because in real systems, the smallest events are at the same level than the measurement noise, meaning that most of the energy is released at these small scales.

If we use the ten-times limits defined in the previous analysis, the T\textsubscript{M8} limits correspond to 7 months and 81 years (Fig. 3f), which are 11,3 times smaller and 12,3 times larger with respect to the real earthquake value of T\textsubscript{M8}=6.6 years in Japan.

Figure 3f shows the differences in b values in cumulative vs. non cumulative distributions. For large b values they are negligible, and only for b<0.5, corresponding to $\tau$<1.33, we see noticeable variations between the two distributions. In the earthquake-like regime of 1.5 $\leq$ $\tau$ < 2.0, both distributions are very similar.

\begin{table*}[t]
  \caption{Dependence of energy release and M8 earthquake statistics on the exponent value $\tau$.}
  \label{tab:exponent_dependence}
  \centering
  \begin{ruledtabular}
  \begin{tabular}{cccc}
  $\tau$ & $\langle E \rangle/\langle E_{\tau=1.67}\rangle$ & $E(M8)/E_{\mathrm{tot}}$ & $T_{M8}$ \\
  0.8 & 5.4$\times$ 10\textsuperscript{7} & 98\% & 3 seconds \\
  0.9 & 3.2$\times$ 10\textsuperscript{7} & 97\% & 4 seconds \\
  1.0 & 1.1$\times$ 10\textsuperscript{7} & 96\% & 13 seconds \\
  1.1 & 1.8$\times$ 10\textsuperscript{6} & 95\% & 1.5 minutes \\
  1.2 & 1.8$\times$ 10\textsuperscript{5} & 93\% & 14 minutes \\
  1.3 & 1.4$\times$ 10\textsuperscript{4} & 91\% & 3 hours \\
  1.4 & 1.1$\times$ 10\textsuperscript{3} & 87\% & 40 hours \\
  \hline
  1.5 & 82 & 82\% & 24 days \\
  \hline
  \textbf{1.58} & \textbf{10} & \textbf{75\%} & \textbf{7 months} \\
  1.6 & 6 & 74\% & 1 year \\
  \textbf{1.67} & \textbf{1} & \textbf{67\%} & \textbf{6.6 years} \\
  1.7 & 0.47 & 63\% & 15 years \\
  \textbf{1.76} & \textbf{0.1} & \textbf{54\%} & \textbf{81 years} \\
  \hline
  1.8 & 4.0$\times$ 10\textsuperscript{-2} & 48\% & 232 years \\
  1.9 & 4.3$\times$ 10\textsuperscript{-3} & 28\% & 3611 years \\
  2.0 & 8.3$\times$ 10\textsuperscript{-4} & 10\% & 56490 years \\
  \hline
  2.1 & 3.3$\times$ 10\textsuperscript{-4} & 2\% & 8.9$\times$ 10\textsuperscript{5} years \\
  2.2 & 2.1$\times$ 10\textsuperscript{-4} & 0.1\% & 1.4$\times$ 10\textsuperscript{7} years \\
  \end{tabular}
  \end{ruledtabular}
\end{table*}
\section{Discussion}
All the previous analyses were carried out under the assumption of constant activity (N\textsubscript{tot}=const.). However, alternative scenarios can also be considered. Here, we focus on the opposite case, in which the total released energy is kept constant (E\textsubscript{tot}=const.) and the annual number of events, N\textsubscript{tot}, becomes the variable of interest.

The energy distributions obtained for different values of $\tau$ (Fig. 4a) differ from those shown in Fig. 3a only by normalization factors. Under the condition E\textsubscript{tot}=const., the curves intersect at high energy values. Nevertheless, when considered individually, all distributions appear qualitatively similar. Independently of the normalization condition --whether constant activity or constant energy release-- one might therefore be tempted to interpret all of them as earthquake-like dynamics, even for $\tau$<1.0, a regime that has indeed been reported in fracture experiments \cite{ref16}.

Converting energies into magnitudes (Figs. 3b and 4b) allows for a more direct comparison with real earthquake statistics. In this representation, it becomes clear that for $\tau$<1.0 small-magnitude events are less frequent than catastrophic ones, a behavior that is in strong contrast with observed earthquake catalogs.

For $\tau$=1.0, where all magnitudes are equiprobable, the activity is extremely low, with N\textsubscript{tot} $\sim$ 2 events per year (Fig. 4c). In this regime, the system spends most of its time accumulating energy, and events occur only rarely. This situation does not correspond to the scale-invariant dynamics commonly observed in experiments, which could otherwise be misinterpreted as earthquake-like behavior. More generally, for $\tau$ < 1.5, the dynamics is characterized by reduced activity and long periods of energy accumulation without avalanches. In both experimental systems and real seismic data exhibiting scale invariance, the number of events is typically very large. For this reason, we consider the condition of constant activity (N\textsubscript{tot}=const.) to be more appropriate for describing and comparing different scale-invariant scenarios as the exponent $\tau$ deviates from 1.67.

Table 1 summarizes the energy-related quantities and the statistics of catastrophic M8 events. In doing so, it provides a quantitative measure of how closely a given scale-invariant dynamics resembles a real earthquake scenario, allowing one to classify the dynamics into the following qualitatively different regimes:

(1) 1.5 $\leq$ $\tau$ < 2.0: Earthquake-like behavior, in the broad sense of a regime characterized by a relatively low input energy rate E\textsubscript{in} leading to scale-invariant energy-release events, in which rare, extremely large events dominate the total energy budget E\textsubscript{tot}. This corresponds to the commonly assumed behavior of faults, where earthquake statistics --with many small events and a few large ones following a power-law distribution-- are taken as a generic illustration of scale-invariant systems. In this view, varying $\tau$ merely adjusts the relative proportion of small to large events while preserving the overall picture.

However, since both the energy ratios and the frequency of M8 events change abruptly as $\tau$ departs from 1.67, we advocate for a stricter notion of proximity to real earthquake behavior. Imposing that the difference in released energy remains within one order of magnitude of that of real seismicity restricts the relevant range to 1.58 $\leq$ $\tau$ $\leq$ 1.76.

(2) For $\tau$ < 1.5, the condition of a relatively low input energy rate E\textsubscript{in} no longer holds. The increased frequency of large-energy events requires a substantial energy supply, which cannot be sustained by slow accumulation alone. In fracture experiments, where $\tau$ = 0.96 has been reported \cite{ref16}, this apparent excess of energy may instead be stored in the elastic bulk of the system.

Note that for $\tau$ < 1, large-magnitude events become more frequent than small ones.

(3) For $\tau$ > 2, the smallest events dominate the total energy budget E\textsubscript{tot}. In most real datasets, such events lie close to the noise level of measurements, making it difficult to reliably estimate the true energy release of the system. This raises fundamental questions about how such systems should be interpreted or modeled. In the context of earthquakes, a value such as $\tau$ = 2.1 would imply that M8 events occur approximately once every million years (instead of once per decade), meaning that a region like Japan would effectively never experience such large earthquakes.

\begin{acknowledgments}
We thank François Pétrélis for clarifying discussions about the Seismic moment. This work was supported by the ANR grant ANR-22-CE30-0046 and the CONYCIT cooperation project MEC80180053.

\end{acknowledgments}

\section*{Author contributions}
O.R designed the research, K.D. and O.R. performed the analysis. K.D., G.V. and O.R discussed the analysis and wrote the manuscript.

\section*{Competing interests}
The authors declare no competing interest.

\section*{Data availability}
\url{https://figshare.com/s/a405ceef78fe379415b7}

\end{document}